\documentclass[prl,twocolumn,showpacs,aps]{revtex4}
\input epsf

\usepackage{amsmath}

\begin{document}

\title{Competition in Social Networks: Emergence of a Scale-free \\
Leadership Structure and Collective Efficiency}

\author
{M. Anghel,$^{1}$ Zolt\'an Toroczkai,$^{1}$\footnote{
Corresponding author: toro@lanl.gov} Kevin E. Bassler,$^{2}$
G. Korniss$^{3}$}
\affiliation{$^{1}$Center for Nonlinear Studies and
Complex Systems Group,  Theoretical Division,
Los Alamos National Laboratory, MS B213 Los Alamos, NM 87545, USA
\\
$^{2}$Department of Physics, 617 Science and
Research bld. I, University of Houston,
Houston, TX, 77204-5005, USA
\\
$^{3}$Department of Physics, Applied Physics, and
Astronomy, Rensselaer  Polytechnic Institute, 110 8$^{th}$ Street, Troy,
NY 12180-3590, USA}

\date{\today}

\begin{abstract}
Using the minority game as a model for competition dynamics, we
investigate the effects of inter-agent communications on the
global evolution of the dynamics of a society characterized by
competition for limited resources. The agents communicate across
a social network with small-world character that forms the static
substrate of a second network, the influence network, which is
dynamically coupled to the evolution of the game. The influence
network is a directed network, defined by the inter-agent
communication
links on the substrate along which communicated information is
acted upon. We show that the influence network spontaneously
develops hubs with a broad distribution of in-degrees, defining a
robust leadership structure that is scale-free. Furthermore, in
realistic parameter ranges, facilitated by information exchange on
the network, agents can generate a high degree of cooperation making
the collective almost maximally efficient.
\end{abstract}


\maketitle

In a competitive environment with seriously limited resources,
an individual will be able to make the most gains, if he avoids
the crowds, and finds strategies that places him into the
distinguished class of the elites, or of the ``few''.
Even though this class forms a {\em minority group} when compared
to the whole agent society, it can largely influence the dynamics
of the entire society for the simple reason that the elites hold
the best strategies in the given situation, and thus they
become key target nodes for others to communicate with, and follow.
For our purposes, an agent is a leader if at least one agent is
following, and thus acting on his advice. The influence of a
leader is measured by the number of followers he has. Agents who
are not leaders are simply coined ``followers''. However, leaders
can follow other leaders, thereby creating a leadership structure.
Certainly, the leadership structure, and even which particular
agents are leaders at all, is often very dynamic (mostly because
the success of a certain strategy is determined by the context of
the strategies used by the other agents).

One of the most ubiquitous mechanisms guiding people in deciding
whom, or what to follow is reinforcement learning\cite{KLM96},
which is a mechanism for statistical inference created through
repeated interactions with the environment. For example, in
iterated situations/games, it can be argued that we all monitor
our social circle, and ``score'' our acquaintances, including
ourselves, based on past performance (success measure).
We then take more seriously, and often follow those with a
higher score (success rate)\cite{SP00}.

In order to study the scenario described above, in this Letter we
use a well known multi-agent model of competition, the Minority
Game\cite{CZ97,SMR99,JHH99} (MG), which we modify to include
inter-agent communications/influences across a social network.
The two main questions we address here are:
1) What type of leadership structure is generated? and 2) Can
the effects of inter-agent communications aggregate up to the
level of  the collective and affect its behavior?

The original MG is an abstraction of a market played by agents
with bounded rationality, inspired by the El Farol bar problem
introduced by Brian W. Arthur\cite{A94}.
In this iterated game, at every step, $N$ agents must choose
between two different options, symbolized by A and B, e.g.
``buy'' and ``sell''. Only agents in the minority group get
a reward. The agents have access to global information,
which is the identity of the minority group for the past
$m$ rounds. Each agent bases his choice on a set of $S$
strategies available to them. A strategy, which is an agent's
`way of thinking', is a prediction \cite{A94} for outcome A or
B, in response to {\em all} possible histories of length $m$.
The strategies are distributed randomly among agents, and
thus in general each agent has different set of $S$ strategies.
They make their next choice in the game using reinforcement
learning: every agent keeps a score for each of the $S$ strategies
which he then increments by one each round if that strategy
correctly predicted the minority outcome (regardless of usage).
The strategy used to make the new choice is the one with the
best score up to that time. If two, or more strategies share
the best score, then one of those strategies is picked randomly.
Previously, the effects of local information in the MG were studied
both with reinforcement learning type \cite{PBC00}
and non-reinforcement learning type \cite{EZ00,KSB00,S00}
of agent communication mechanisms on Kauffman networks
\cite{K93}, and with non-reinforcement learning type of
mechanisms on linear chains \cite{KSB00,S00}.


In our model, a social network of agents is described by a
graph with vertices representing the agents, and edges
representing acquaintanceship between pairs of agents. This
network of acquaintances forms the substrate network
({\bf G}), or skeleton for inter-agent communications
\cite{SP00,PBC00,EZ00,KSB00,S00,NM92}. An edge $ab$ in ${\bf G}$
means that agents $a$ and $b$ may  exchange game-relevant
information. However, it does not indicate whether the exchanges
influence the action by any of the involved agents. That
information is modeled by a second network, the influence
network $({\bf F})$, which is a directed subset of
${\bf G}$, and in which an edge $ab$, pointing from $a$ to $b$,
means that agent $a$ {\em acts} on the advice
of agent $b$ when deciding the minority choice.
In the competitive environment of the stock market, Kullman,
Kert\'{e}sz and Kaski, by studying time-dependent
cross-correlations have recently shown the existence of such a
directed network of influence among companies(${\bf F}$)
based on data taken from the New York Stock Exchange \cite{KKK02}.
We do not, in general, know the precise topology of the social
networks. However, it is known that social networks have a
small-world character \cite{WS98,W99,JGN01}.  Here we take
${\bf G}$ to be an Erd\H{o}s - R\'enyi (ER) random graph with link
probability $p$. An ER random graph shows the small world effect,
since the diameter of the graph increases only logarithmically
with the number of vertices \cite{N00} and the nodes also have a
well defined average degree, $pN$, which results from cognitive
limitation \cite{JGN01}. Studies using other types of network
topologies, which are more suited to describe social networks (one
drawback of ER is its low clustering coefficient \cite{WS98}) will
be presented in future publications.   Just as in the original MG,
in our model, in order to make his next decision, each agent
uses his best performing strategy to predict what the next minority
choice will be. However, he does not necessarily act on
that prediction.
Instead, the prediction simply constitutes the agent's opinion,
which he then shares with all his first neighbors on the
substrate network ${\bf G}$. This is done by all agents
simultaneously, and thus every agent obtains as information the
predictions of all their first neighbors. Each agent then uses this
information to make their final choice, via reinforcement learning:
they keep scores of the prediction performance of all their
first neighbors and themselves,
and update the scores after every round by incrementing the
scores of the agents whose prediction was correct. Each agent then
acts on the prediction/opinion of the neighboring agent with
the highest score. Of course, if they have a higher score than any
of their neighbors, then they act on their own prediction.

\begin{figure}
\centerline{\epsfxsize = 3.4 in \epsfbox{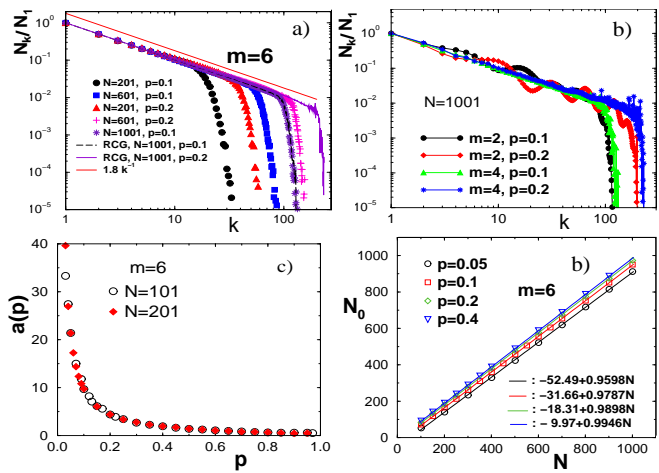}}
\vspace*{-0.2truecm}
\caption{Leaders and followers.
a) and b) show the average of the number
of leaders with $k$ followers normalized by the average
number of leaders with exactly one follower  $N_1$.
The symbols correspond to varying system sizes
and link probabilities, $p=0.1$ and $p=0.2$,
respectively, while the dashed and
thin continuous lines correspond to the same quantity
for the Random Choice
Game on the ER substrate. Next to the curves, the
thick continuous line has a slope of -1.
b) shows the same
quantity for small memories, $m=2$ and $m=4$ with $S=2$ for
$p=0.1$ and $p=0.2$. The curves oscillate
around the same $1/k$
law. For all curves in  a) and b) the averages
were taken over 17
runs, which was sufficient, due to the
strong self-averaging
property of the quantities. 
c) shows that $a(p)\equiv N_1$ with
good approximation is independent on the system size $N$. d)
represents the number of followers as a function of the system
size $N$. Both for a) and b), $m=6$ and  $S=2$.}\label{fig1}
\end{figure}

The game is initialized by fixing at random $S$ strategies
for each agent, an arbitrary initial history string, and a
fixed instance of the substrate network ${\bf G}$.
After many iterations, the game evolution becomes insensitive to
the particular initial history string.
However, it may remain sensitive to the quenched disorders in
the strategy space of the $NS$ strategies that are used, and in
the quenched disorder associated with the particular social
network chosen. Thus, there are four relevant parameters in this
game: $N$, $S$, $m$, and $p \in [0,1]$. Of course, in
reality the substrate network can also change (we  make new
friends and others fade away). However, we assume its dynamics
to be much slower than that of ${\bf F}$, and therefore it is
neglected here. As defined previously, an agent  $i$  is a leader
if it has at least one follower,  $j$, and thus agent  $j$ follows
through action what agent $i$ suggests. For this to happen, $i$
has to have the largest prediction score among the acquaintances
of $j$, which are defined as the $k_j$ edges $j$ has
in ${\bf G}$. In an ER graph, the number of $k_j$  links has a
Poisson distribution with an average
value at $\lambda=pN$, and an exponential tail.
An agent $j$ will follow only one agent's opinion
to decide his action, and thus its number of out-links is
always one, $k_j^{(out)} = 1$.
However, the number of in-links for agent $j$, $k_j^{(in)}$,
can be any number between $0$ and $k_j$, according
to the number of agents acting on his advice.
Fig.\ref{fig1} shows the in-degree distribution for various
numbers of agents $N$, network connectivity $p$,
and memory length $m$. The first striking observation
from Fig.\ref{fig1}a) is that over a wide range
of parameters the in-link distribution is described
by a power-law with a sharp cut-off. Thus,
the average number of leaders with $k$ followers, $N_k$,
is a scale-free distribution \cite{AB02}. This
happens in spite of the fact that the substrate
network, which is an ER graph is not a scale
free network, and therefore it was not introduced
{\em a priori} into the underlying structure. The
scale-free character of the influence network
${\bf F}$ is selected for by the reinforcement learning
nature of the agent-agent interaction rules.
The fact that a broad scale-free structure is
selected on the back of a Poisson distributed network,
seriously limits the size of the leadership. Indeed,
Fig.\ref{fig1}d), which shows the non-leaders, or followers,
expresses this fact: the pure followers constitute
over 90\% of the population for the cases
presented in Fig.\ref{fig1}a).

Plotting $N_k / N_1$, all the curves can be collapsed in the
scaling regime up to their cut-offs,
indicating that $N_k(N,m;p) \propto k^{-\beta} N_1(N,m;p)$.
The power of the decay, $\beta$ is very
close to unity, which means that $kN_k$ is independent of $k$
and the other parameters in the scaling regime.
Since $k$ is the influence of a leader with $k$ followers,
$kN_k$  represents the total influence
of the $k$-th layer in the leadership hierarchy. The above
observation therefore means that
all layers of the hierarchy are equally influential;
influence is evenly distributed among
all levels of the leadership hierarchy. This result is
robust, and insensitive to the particular
parameters, even in the low $m$ (memory) regime. Here, however,
oscillations build up around the
$1/k$ behavior which still serves as a backbone for the
leadership structure, but it becomes less
obvious as $m$ is decreased, see Fig.\ref{fig1}b).
Another important observation is that $N_1(N,m;p)$ depends
strongly only on $p$ and not on $N$ or $m$,
thus $N_1(N,m;p) = a(p)$, as shown in Fig. 1c).
Therefore, we have
\begin{equation}
N_k(N,m;p) = k^{-\beta} a(p) f_k(N,m;p).
\end{equation}
The fact that $N_1(N,m;p)$ is virtually independent of
$N$, means that if the number of agents is increased, the
leadership structure and size {\em in the scaling regime} will
not change! What changes though, is the number of the
``sheep'' or followers, which is $N_0$. It will
grow in proportion to $N$, as seen in Fig.\ref{fig1}d).
Also, the cut-off at the high-$k$ end of the
distribution will occur at larger $k$ as $N$ is increased.
The deviation of the
function $f_k(N,m;p)$
from a constant accounts for the fluctuations in the
leadership structure which vanish (the
fluctuations) with increasing $m$. This is due to the fact
that the strategy space suffers a combinatorial
explosion as $m$ is increased (there are in total
$2^{2^m}$ strategies), and the agents' strategies
therefore become highly uncorrelated \cite{CZ97,SMR99,JHH99}.

This suggests that the results
for large $m$ can be reproduced if the agents simply
play a Random Choice Game (
RCG) on the network.
In a RCG, agents do not use strategies, but instead just
toss a coin when making
 predictions. Indeed,
Fig. 1a) shows that the RCG on the ER network produces
the same scale - free backbone of the
leadership structure. Thus, in our model the closeness to
the scale-free backbone is determined
by the level of mutual de-correlation of  agents' strategies.
This is to say that increased
trait diversity (strategy space) leads to stable scale-free
leadership structure.

Although the leadership structure is stable for large $m$,
the position of an individual agent in the leadership
hierarchy is not. By computing the time correlations present
in the number of in-links we can show that the average
lifetime of an agent in a particular leadership position is
short for large $m$, as detailed in Ref. \cite{proc}.
In contrast, at low $m$ values, leaders become frozen
in their positions. In other words, in the low $m$ regime,
where trait diversity is small, as in a dictatorship, where agents'
action space is severely limited, leaders ``live''
longer in their positions.


Next, we briefly study the global performance of
the collective on the network. Consider
choice A as the reference option, and denote by
$A(t)$  the attendance, or the number of agents
choosing option  A  at time $t$.
One of the most frequently used measures for a
``world utility'' function for the collective
\cite{WT02} is the variance $\sigma$ of the fluctuations
in the time series of $A(t)$. In the language of economics,
it is the volatility of the market, and from a systems
design point of view \cite{WT02} it is the quantity
that we ultimately want to minimize.

\begin{figure}
\centerline{\epsfxsize = 3.4 in \epsfbox{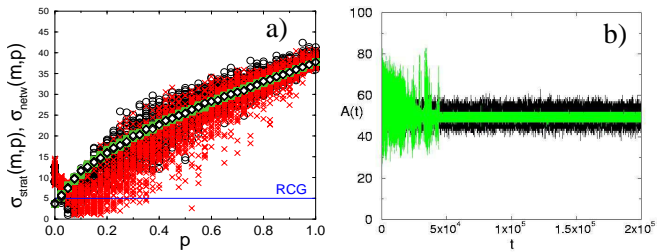}}
\vspace*{-0.3truecm}
\caption{Collective efficiency.  a) shows the time-averaged
volatility (over $5\times 10^5$ steps) of the market
as a function of the substrate network
connectivity parameter, $p$. The empty  circles
($m=2$) and the solid squares
($m=8$) are obtained by fixing the strategy
space disorder and taking randomly 50
network samples, while the crosses ($m=2$)
and the diamonds ($m=8$) are obtained with the network
space disorder fixed for 50 strategy disorders.  Here
$S=2$ and $N=101$.  b) shows a sample time series in 
(green/gray)
for one of the low lying points in a) at $p = 0.1$, $m=2$.
The black time series corresponds to a run for the
ordinary MG at minimum volatility
which is at $m=6$, $S=2$.
The black curve has a variance of 2.36, while the green/gray
has a variance of 1.07.} \label{fig2}
\end{figure}

As mentioned before, this game has two types
of quenched disorder embedded into it. A natural
question then is if one can find/evolve networks
that achieve zero, or almost zero volatility given a
group and their strategies, or, alternatively, if one
can find strategies that achieve zero, or near zero
volatility, given a particular substrate network.
To answer this question, we performed simple random
searches in one of the quenched disorder spaces (network
or strategy) keeping the other quench disorder fixed
(strategy or network). An example with
$m=2$ and $m=8$  is displayed in Fig.\ref{fig2}a)
as a function of connectivity $p$. The first conclusion is
that overall, the collective does  worse with
``smart agents'' (large $m$) on highly connected networks
if they exchange information about their strategies.
However, in the low $m$ regime ($m=2$), the
system efficiency can improve not only beyond that
of the standard MG,  but also beyond that of
the RCG without network (blue line with
$\sigma_{RCG} = 0.5 \sqrt{N}$), and even beyond the
standard MG's best performance (which is at a different
value of $m=6$ for these parameters).  Thus, a networked,
low trait diversity system can be more effective as a
collective, than a sophisticated group.  Note that
the optimal $p$ values are still much larger than
the critical value for the giant component in
the ER network, which is $1/N$, and thus we need well
connected single component graphs in order
to observe the collective efficiency emerge from the
agent-agent  interactions. However, the
optimal values are actually in the realistic range
for social networks, giving for the average number of contacts
$\lambda = pN \simeq 10 - 20$.  If $N$ is varied
the optimum range for $p$ shifts such that
optimum value of $pN$ remains constant.
Fig.\ref{fig2}b) shows a sample time-series from the
optimal connectivity region. Notice the low
volatility compared to the best performance
of the MG (in the background). In the
standard MG the variations in $\sigma$ at
the best performance point are  low, and even an extended
search (500 samples) in the strategy
disorder space could not generate  $\sigma$-s lower than 2.0,
while in contrast, time series such as the red one in
Fig.\ref{fig2}b) are easily generated within 50
random samples in the optimal connectivity region.
This emerging collective efficiency can be
understood in terms of the crowd-anticrowd
description of the MG, as introduced by Johnson, Hart
and Hui \cite{JHH99}. In the MG, low $m$ means
that only a small number of different strategies
are possible, thus many agents are forced to use the
same strategy and thus they behave as a crowd, or a group.
This grouping effect generates the large volatility in
the ordinary  MG. When the game is played
on a network, however, an agent, even if it shares the
same strategy as the others in a large
group, now has the  possibility to listen to some
other agents, and possibly even from other groups.
Thus, it is no longer forced to behave the same way
as its own group, thereby breaking the grouping
behavior. If, however,  $p$ is too large, there is a
grouping behavior appearing due to the  network,
because an agent will have too many followers if his
score is the highest, creating a group on
the network. The two crowding effects compete and a
balance between them is reached in the optimum
connectivity region.


In summary, we have shown that the evolution of
multi-agent games can strongly depend
on the nature of the agent's information resources,
including local information gathered
on the social network, a network whose structure
in turn is influenced by the fate of the game itself.
In our study, we allowed for this dynamic coupling between the
game and the network by using reinforcement learning
as an ubiquitous mechanism for
inter-agent communications. Our observations are:
1) if reinforcement learning is used, a scale-free
leadership structure can be created,
even on the backbone of non-scale free networks;
2) in low trait diversity collectives, enhanced
collective efficiency may appear, making
this effect worthwhile for systems design studies \cite{WT02}.

Z.T. and M.A. are supported by the Department of Energy
under contract W-7405-ENG-36,
K.E.B. is supported by the NSF-DMR through 0074613,
and the  Alfred P. Sloan Foundation.
G.K. is supported by the NSF-DMR through DMR-0113049
and the Research Corporation through RI0761.

\end{document}